\documentclass[12pt]{article}
\usepackage{graphicx} % Required for inserting images
\usepackage{geometry}
\usepackage{amsmath}
\usepackage{comment}
\usepackage{amsfonts}
\usepackage{amssymb}
\usepackage{booktabs} 
\usepackage{subcaption}
\usepackage{xcolor}
\usepackage{url}
\usepackage{float}
\usepackage{caption}
\usepackage{booktabs}
\usepackage{relsize}
\usepackage[siunitx]{circuitikz}
\usepackage{amsmath}

\title{A method for the numerical analysis of hybrid lumped-distributed superconducting  quantum circuits}
\author{
\begin{tabular}{c}
Simona Zaccaria\textsuperscript{1,2} \quad Antonio Gnudi\textsuperscript{1,2} \\
\\
\textsuperscript{1}Department of Electrical, Electronic and Information Engineering \\
``Guglielmo Marconi'', University of Bologna \\
Viale del Risorgimento 2, 40136 Bologna, Italy \\
\\
\textsuperscript{2}ARCES - Advanced Research Center on Electronic Systems \\
University of Bologna \\
Via Toffano 2/2, 40125 Bologna, Italy
\end{tabular}
}
\date{}

\begin{document}

\maketitle
\begin{abstract}
We present a method for the numerical analysis of superconducting quantum circuits combining lumped elements, either linear or non-linear (i.e.~Josephson junctions), and distributed coplanar waveguide (CPW) structures. CPW transmission lines and multiline couplers are directly modeled without discretizing them into lumped-element equivalents, and
the circuit Hamiltonian parameters are extracted by using the energy participation ratio (EPR)
method. This approach enables fast and accurate extraction of mode frequencies,
anharmonicities, cross-Kerr interactions, and Purcell decay rates without relying on
full electromagnetic simulations, while naturally accounting for higher-order modes
of distributed components. 

We have implemented the proposed method in a Python framework, QuLTRA (Quantum hybrid Lumped and TRansmission lines circuits
Analyzer), which we have used to validate the approach against full electromagnetic simulations (Ansys HFSS, pyEPR), existing circuit-analysis tools (QuCAT), and designs reported in the literature. The comparisons show excellent agreement with orders-of-magnitude reductions in computational time relative to full-wave solvers. We demonstrate applications including Purcell-protected readout, multimode ultra-strong coupling, and multiplexed qubit readout, illustrating how the method can support fast and reliable early-stage circuit design.
\end{abstract}
\section{Introduction}
Superconducting quantum circuits are one of the most promising platforms for quantum computation and have experienced significant development over the past two decades \cite{Blais_2021, Krantz_2019}. One of the reasons for this rapid growth can be found in the high flexibility offered by this platform as compared to alternative technologies, due to the circuit nature of the implementation, where lithographically defined artificial atoms can be controlled, coupled, and read out in many efficient ways. 

The design of superconducting quantum circuits is in general a nontrivial task.  
Starting from a Hamiltonian model, which is optimized for a particular purpose, one seeks a circuit whose topology and component parameters have the appropriate values for the resulting circuit Hamiltonian to match the desired one. The aim of the circuit design phase is to determine these values, including the ones defining the interconnections between different components, which is one of the reasons why this phase is often quite complex. The task is very time-consuming because Hamiltonian parameters can be calculated from a circuit design, but the opposite is not possible. Therefore, circuit designers need to make an initial guess, extract the Hamiltonian and verify it, modify the circuit if necessary, and repeat the process iteratively. The availability of efficient tools for the circuit simulation and extraction of quantum parameters is therefore quite important.

Several open-source software packages have been developed over the years for the above purpose. Some of them adopt a quantization procedure in the basis of eigenmodes of the linearized circuit (normal modes), and then approximately account for non-linear effects due to Josephson junctions in a perturbative way to extract important Hamiltonian parameters, such as cross-Kerr and self-Kerr (anharmonicity) coefficients. This method is particularly suitable for weakly anharmonic circuits, which are at the basis of recent superconducting implementations of quantum processor prototypes \cite{Koch_2007}.
Two distinct strategies have been adopted for the determination of the linearized circuit modes. In the first one, a full finite-element electromagnetic eigenmode analysis is carried out. This is, for example, the methodology adopted by the popular Qiskit Metal + Ansys HFSS + pyEPR \cite{QMetal, ansia, minev_epr} combination of tools. This type of analysis is quite accurate, but computationally highly demanding and practically viable only for relatively simple circuit layouts.
On the other hand, other tools such as, for example, QuCAT \cite{Gely_2020} adopt an impedance response matrix approach to determine the normal modes, calculating the impedances seen by each non-linear element of the circuit, but handle only lumped components, that is networks of resistances, inductances, capacitances, and junctions.

A different class of tools, e.g. SQcircuit \cite{SQcircuit} and CircuitQ \cite{CircuitQ}, perform the quantization in a mixed harmonic, flux or charge basis to improve flexibility and reduce the computational requirement for diagonalization. However, to our knowledge, such tools handle only lumped components.
All superconducting quantum circuits, however, contain parts, such as transmission-line based resonators, couplers and interconnections, that cannot be naturally described as lumped elements, unless one discretizes them into a suitable number of inductor and capacitor sections.

In this work we present a general analysis and quantization method that naturally incorporates both lumped and distributed components. The procedure is based on representing the linearized circuit as a network of interconnected multi-port components described by their admittance matrices. This framework allows distributed coplanar waveguide (CPW) structures to be treated directly through analytical expressions, without discretizing them into LC ladders, while still enabling the extraction of normal modes, and energy participation ratios (EPR).

To demonstrate and validate the method, we have implementated it in the Python package QuLTRA (Quantum hybrid Lumped and TRansmission lines circuits Analyzer). In its present version QuLTRA supports only ideal (i.e., lossless and straight) coplanar waveguide (CPW) transmission lines and
CPW couplers within the class of distributed components. However, we stress that the approach is fully general and can accommodate any linear building block for which an admittance matrix is available. Such matrices may be obtained analytically, possibly including empirical corrections, or derived numerically from electromagnetic simulations, and can be incorporated into a component library to extend the range of supported structures.

Just to give an idea, a circuit containing a CPW quarter-wavelength ($\lambda/4$) resonator can be directly analyzed once the physical length and characteristic impedance of the CPW line are assigned, eliminating the need of approximating its behavior with a number of parallel LC sections and automatically accounting for higher resonant modes. 
As an additional example concerning CPW couplers, the dissipation rate of a resonator inductively coupled to a readout transmission line can be directly calculated with QuLTRA. 

The main advantage offered by this analysis method, illustrated in its QuLTRA implementation, is that
it enables fast iterations of the guess-and-check process during the early design phase, significantly reducing the need for computationally expensive electromagnetic simulations, which can be employed in a final refinement and verification phase.
An additional feature of QuLTRA is its capability to model a series of identical Josephson junctions through a single equivalent junction, thus reducing the number of nodes and speeding up the computation.

The QuLTRA Python package used to generate the results presented in this paper is publicly available on GitHub \cite{github}, which also includes tutorials covering many of the examples discussed in this work, and archived on Zenodo \cite{zenodo}.

The remaining part of the paper is organized as follows. Section~\ref{sec:methodology} describes the theoretical framework and the numerical procedure adopted by QuLTRA. Section~\ref{sec:components} provides models and implementation details for the supported components. In  Section~\ref{sec:validation} test cases are presented for validating the implementation against electromagnetic simulations, other existing software tools, and literature results. In Section~\ref{sec:examples} some application examples are discussed, and, finally, conclusions are drawn in Section~\ref{sec:conclusions}.

\section{Methodology overview}
\label{sec:methodology}
The theoretical background of the quantization method of Josephson junction circuits adopted in this work can be found in \cite{minev_epr} \cite{JJ_Filter} and relative supplementary information. The main concepts and equations are reported hereafter for an easier understanding of the rest of the paper.

The Hamiltonian of a circuit containing Josephson junctions can be written as the sum of two terms: the first one is the linear part, including also the linear contribution of the junctions, the second one accounts for the nonlinear behavior of the junctions. In the following, $M$ is the number of relevant eigenmodes of the linearized circuit (in principle in a circuit with distributed resonators there are infinite modes, but only a few of them are relevant for practical applications), and $\omega_m$ is the angular frequency of the $m$-th mode. Besides, for generality, it is assumed that the circuit contains $N_{JJ}$ arrays of Josephson junctions, the $i$-th of which is formed by $N_i$ identical junctions in series, with a total linear inductance $L_i$ (hence, $N_i = 1$ for a single junction). The Hamiltonian then reads
\begin{equation}
    \hat{H}=\sum_{m=1}^M{\hbar \omega_m \hat{a}_m^\dagger \hat{a}_m}-\sum_{i=1}^{N_{JJ}} N_i^2 E_i [\cos(\hat{\phi}_i/N_i)+\hat{\phi}_i^2/(2N_i^2)]
    \label{Ham1}
\end{equation}
where $E_i=(\frac{\hbar}{2e})^2\frac{1}{L_i}$, $\hat{a}_m^\dagger$ and $\hat{a}_m$ are the creation and annihilation operators for mode $m$, respectively, and $\hat{\phi}_i$ is the reduced flux operator across the $i$-th array of junctions, which can be written as 
\begin{equation}
\phi_{i} = \sum_{m=1}^M \phi_{m,i}(a_{m}^\dagger +a_{m})
\label{phi_i}
\end{equation}
$\phi_{m, i}$ being the dimensionless, real-valued, quantum zero-point fluctuation of the $i$-th reduced flux in mode $m$. It turns out that the Hamiltonian is completely determined once the linear mode eigenfrequencies $\omega_m$ and the flux zero-point fluctuations $\phi_{m, i}$
have been calculated. The latter can be written in terms of the energy-participation ratio (EPR) $p_{m,i}$ of the $i$-th array in mode $m$ as \cite{minev_epr} 
\begin{equation}
 \phi_{m,i}^2 = p_{m,i}\frac{\hbar \omega_{m}}{2E_{i}} 
\label{pmj}
\end{equation}
where 
\begin{equation}
    p_{m,i}=\frac{\text{linear inductive energy stored in the $i$-th array in mode $m$}}{\text{total inductive energy stored in mode $m$}}
    \label{pmi}
\end{equation}
The specific expressions for $p_{m,i}$ are component-type dependent and  will be illustrated in Section 3.

Expanding in \eqref{Ham1} the cosine to fourth order and using \eqref{phi_i}, one obtains
\begin{equation}
     \hat{H}\simeq \sum_{m=1}^M{\hbar \omega_{m} \hat{a}_{m}^\dagger \hat{a}_{m}}-\sum_{i=1}^{N_{JJ}} \frac{E_{i}}{24N_{i}^2}\left[\sum_{m=1}^M{\phi_{m,i}(a_{m}^\dagger +a_{m})}\right]^4
     \label{Hamiltonian}
\end{equation}
which can be further manipulated replacing \eqref{pmj} and applying the rotating-wave approximation, to obtain the effective Hamiltonian
\begin{equation}
    \hat{H} \simeq \sum_{m=1}^M  \left[ \hbar(\omega_{m}- \Delta_{m}) \hat{a}_{m}^\dagger \hat{a}_{m} - \frac{1}{2} \hbar \alpha_{m} \hat{a}_{m}^{\dagger 2} \hat{a}_{m}^2 \right]\\ 
- \sum_{n \ne m} \frac{1}{2} \hbar \chi_{mn} \hat{a}_{m}^\dagger \hat{a}_{m} \hat{a}_{n}^\dagger \hat{a}_{n}
\label{final_hamiltonian}
\end{equation}
where $\Delta_{m}$ and $\alpha_{m}$ are the effective Lamb shift and anharmonicity of mode $m$, respectively, and $\chi_{mn}$ is the cross-Kerr  coefficient between modes $m$ and $n$, given by
\begin{subequations}
\label{NLpar}    
\begin{align}
        \begin{split}
        \label{chi_a}
        \chi_{mn}& = \sum_{i=1}^{N_{JJ}} \frac{1}{\hbar} \frac{E_{i}}{N_{i}^2} \phi_{m,i}^2 \phi_{n,i}^2 = \sum_{i=1}^{N_{JJ}} \frac{\hbar \, \omega_{m} \omega_{n}}{4 E_{i}} \frac{p_{m,i} p_{n,i}}{N_{i}^2} \end{split}\\
        \label{chi_b}
        \Delta_{m}&=\frac{1}{2}\sum_{n} \chi_{mn}\\
        \label{chi_c}
        \alpha_{m}&=\frac{\chi_{mm}}{2}
\end{align}
\end{subequations}
From \eqref{final_hamiltonian} and \eqref{NLpar} it is clear that, in order to fully determine the Hamiltonian, one needs to know the mode frequencies and the cross-Kerr coefficients. QuLTRA numerically computes the mode frequencies and then derives the cross-Kerr matrix by evaluating $p_{m,j}$ for each Josephson junction array.

\subsection{Calculation of the linear eigenmode frequencies}
As mentioned in the introduction, the method applies to any Josephson circuit that, after replacing each junction array with its equivalent linear inductance, $L_{i}$, can be modeled as a network of linear multi-port components, each described by an admittance matrix that allows to express the port currents as a linear combination of the port voltages \cite{Pozar}.

To analyze the circuit, one node is selected as the reference (ground node), and the voltages of the remaining nodes are defined relative to it. Kirchhoff's current law is then applied to each non-ground node. The resulting set of equations can be written in matrix form as
\begin{equation}
    \mathbf{Y}(z)\, \mathbf{V} = 0
\end{equation}
where $\mathbf{V}$ is the array of the phasors of node voltages, $\mathbf{Y}(z)$ the total admittance matrix of the circuit and $z$ a complex frequency variable. The eigenmode frequencies are the values of $z$ solutions of the equation 
\begin{equation}
\det(\mathbf{Y}(z)) = 0
\label{det}
\end{equation}
The $i$-th solution $z_i$ corresponds to the non-null eigenvector $\mathbf{V}_i$ of node voltages. In general, in a circuit with losses, $z$ is complex and is written as 
$z = \kappa/2 + j\omega$, with $\omega$ the mode frequency and $\kappa$ the mode dissipation rate (or linewidth).
%
%\begin{equation}
%    I_\mathrm{i} = \sum_j Y_\mathrm{ij}(z) V_\mathrm{j}(z).
%\end{equation}
%
In the case of a lossless circuit this reduces to $z = j\omega$.

The following strategy, which has been tested in QuLTRA, is proposed to find the roots of \eqref{det} with $\omega = 2 \pi f$ with frequency $f$ in the range $[f_\mathrm{min}, f_\mathrm{max}]$.
\begin{enumerate}
    \item \textbf{Lossless circuit}. The frequency range $[f_{\min}, f_{\max}]$ is split into a sequence of partially overlapping intervals. 
    %The subinterval width and the overlap between consecutive intervals can be adjusted by the user (the default values are $0.2$\,GHz and $0.1$\,GHz, respectively).
    The starting points of the intervals are spaced by a fixed step. Each interval has a width equal to two steps, which implies that consecutive intervals overlap by exactly one step. By default, the step is set to 0.1 GHz, but it can be adjusted by the user.
    The overlap mitigates the risk of missing roots that lie close to the boundaries of the intervals. For each interval, Brent’s method \cite{brent2002algorithms}, which combines the secant and the bisection root-finding algorithms and is implemented in the \texttt{scipy.optimize} library through the \texttt{brentq} function, is used to find the possible roots within that interval.  
    Since \texttt{brentq} requires a real-valued function with opposite signs at the interval endpoints, depending on the size $n$ of matrix $\mathbf{Y}(z)$, either the real or the imaginary part of the determinant is taken. This is because, being all the terms of the matrix $\mathbf{Y}(z)$ purely imaginary, its determinant is proportional to $j^n$, that is purely real or purely imaginary depending on $n \bmod 2$. To discard false zeros caused by sign changes of the function due to poles, a check is performed by retaining only those found zeros $z_o$ for which $|\det(\mathbf{Y}(z_0))| < 10^{-3}$.
    \item \textbf{Lossy circuit}. The lossies considered here are the ones due to the resistive terminations of the transmission lines. In this case the algorithm is divided in two parts. In the first part, the method described above is used to calculate the zeros of the lossless circuit obtained by short-circuiting all the resistors in the original circuit. Assuming all the resistors have one terminal connected to ground (the most common condition, see the examples in the next sections), the admittance matrix of the lossless circuit is obtained simply by removing the rows and columns corresponding to the off-ground terminals of the resistors. In the second part, each purely imaginary detected root is passed as initial guess to the Newton-Raphson method from the same \texttt{scipy.optimize} library. The latter method is typically fast converging in practical circuits where losses are small and zeros close to the imaginary axis.

\end{enumerate}
\subsection{Calculation of the energy participation ratios}
The EPRs $p_{m,i}$ are calculated using directly the definition \eqref{pmi}.
For each zero $z_m$ found with the above method, QuLTRA computes the corresponding eigenvector $\mathbf{V}_m$ of the node voltages with the \texttt{scipy.linalg.null\_space} function from the SciPy library. From the node voltages, 
the contribution of each component to the total inductive energy is evaluated. The specific expressions are component-type dependent and are derived in the next section for the components implemented so far in QuLTRA.

\section{Component equations}
\label{sec:components}
In its present version, our implementation applies to resistors, capacitors, inductors and Josephson junctions among lumped components, coplanar waveguide (CPW) transmission lines and couplers with straight lines among distributed components. It is worth emphasizing, however, that the approach adopted here is completely general and not limited to these specific components as long as the linearized circuit consists of a network of interconnected multi-port subcircuits, each described by an admittance matrix.

All lumped components are two-terminal elements fully characterized by their admittance. Specifically, the admittance is given by $Y(z) = z\,C$ for capacitors, $Y(z) = 1/R$ for resistors, and $Y(z) = 1/(z\,L)$ for inductors. In the case of Josephson junctions, $L$ represents the equivalent linear inductance of the junction.
For inductors and junctions the inductive energy is calculated as $E_L = \frac{1}{2}L|I|^2$, where the current $I$ through the component is derived from Ohm's law, $I = Y\,V$, where $V$ is the voltage difference between the two terminals. 

The treatment of distributed components, CPW lines and couplers, is more involved and is addressed separately in the two following subsections. 

\subsection{CPW transmission line}
A CPW transmission line consists of a central conducting strip separated by two lateral gaps from two conducting planes lying on the same substrate. The conducting planes can be treated as ground references. Our implementation (QuLTRA) supports the model of straight and lossless CPW lines. From circuit theory of transmission lines \cite{Pozar,Collin}
%
%\begin{equation}
%\frac{dV(x)}{dx} = -j\omega L\, I(x) \qquad \frac{dI(x)}{dx} = -j\omega C\, V(x)
%\end{equation}
%
it turns out that the voltage and current phasors along the spatial coordinate $x$, $V(x)$ and $I(x)$, read 
\begin{subequations}
\begin{align}
    V(x) &= V^+ e^{-jkx} + V^- e^{jkx} \label{uno} \\
    I(x) &= I^+ e^{-jkx} + I^- e^{jkx} \label{due}
\end{align}
\label{phasor_solution}
\end{subequations}
In the above, \(k = \omega/v \) is the wave propagation constant, where $v$
is the wave propagation velocity along the line. The latter can be expressed as $v=1/\sqrt{LC}$ in terms of the inductance and capacitance per unit length $L$ and $C$, respectively, or equivalently as $v=c/\sqrt{\varepsilon_\mathrm{eff}}$, with 
$c$ the velocity of light in vacuum and $\varepsilon_\mathrm{eff}$ the effective relative permittivity of the surrounding medium. For a CPW line on a substrate with relative permittivity $\varepsilon_{r}$, it is $\varepsilon_\mathrm{eff} = (\varepsilon_{r} + 1) / 2$. The coefficients \(V^\pm\) and \(I^\pm\) represent the amplitudes of the forward ($+$) and backward ($-$) traveling voltage and current waves, respectively, which are related by 
\begin{equation}
V^\pm = \pm Z_0 I^\pm
\label{pm}
\end{equation}
where \(Z_0 = \sqrt{L/C}\) is the characteristic impedance of the line.
%
%By substituting one of the telegrapher’s equations into the other, the following one-dimensional Helmholtz equations are obtained:
%\begin{equation}
%\frac{d^2V(x)}{dx^2} + \omega^2LC\,V(x) = 0 \qquad
%\frac{d^2I(x)}{dx^2} + \omega^2LC\,I(x) = 0
%\end{equation}
%

A transmission line of length $\ell$ is a 2-port component, whose admittance matrix is easily calculated using \eqref{phasor_solution}.
%
%To compute the admittance matrix of a CPW segment of length $\ell$, we use the Eq. \ref{phasor_solution}. The method consists in imposing test boundary conditions and evaluating the resulting currents at the endpoints.
%
%First, we set the voltage at port 1 to 1 and at port 2 to 0, i.e., $V(0) = 1$, $V(\ell) = 0$, we determine $V^+$ and $V^-$ and then compute the current. Evaluating $I(0)$ and $I(\ell)$ yields the first column of the admittance matrix. The same procedure is repeated with boundary conditions $V(0) = 0$, $V(\ell) = 1$ to obtain the second column.
%
Its expression, after the replacement $j\omega \rightarrow j\omega+\kappa/2 = z$ is applied, reads 
\begin{equation}
\mathbf{Y}_\mathrm{tml}(z) =
\frac{1}{Z_0 \sinh(z\ell /v)}
\begin{pmatrix}
\cosh(z\ell /v) & -1 \\
-1 & \cosh(z\ell /v)
\end{pmatrix}
\end{equation}
The expression of the inductive energy stored in the CPW line for mode $m$ is derived in a similar way. From the knowledge of the node voltages at the two endpoints for mode $m$, $V(0)$ and $V(\ell)$, the constants $V^\pm$ are calculated from \eqref{uno}, followed by the calculation of $I^\pm$ with \eqref{pm}, thus determining the current profile $I(x)$ in \eqref{due}.
The inductive energy is then given by the integral
\begin{equation}
    E_\mathrm{tml} = \frac{L}{2} \int_0^\ell |I(x)|^2 \, \mathrm{d}x
\end{equation}
which can be evaluated analytically. 

The model is strictly valid if the line is straight, with no bends or curves. Any curvature in the actual layout may introduce slight deviations from the model predictions.

In the following examples the relative permittivity $\varepsilon_r = 11.9$ (silicon) is assumed.

\subsection{CPW coupler}
\begin{figure}
    \centering
    \begin{subfigure}[b]{0.45\textwidth}
        \centering
        \includegraphics[width=\textwidth]{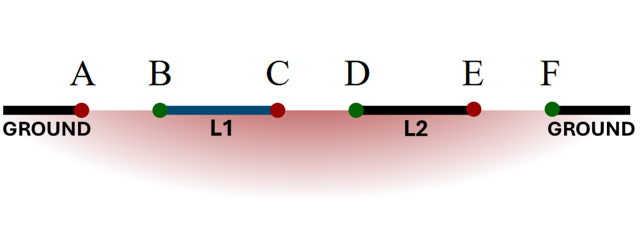}
        \caption{}
        \label{no_ground}
    \end{subfigure}
    \hfill
    \begin{subfigure}[b]{0.45\textwidth}
        \centering
        \includegraphics[width=\textwidth]{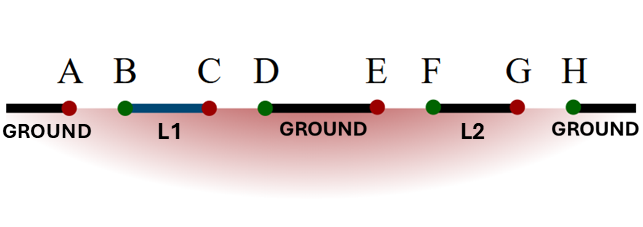}
        \caption{}
        \label{ground}
    \end{subfigure}
    \caption{Cross-section of a multi-line CPW structure formed by two lines (L1 and L2) with (b) and without (a) a grounded central line.}
    \label{cpw_coupler}
\end{figure}

Another structure of interest is the multi-line CPW structure. In this work, we focus on the case of two signal lines with or without a third grounded line in between (Fig.\,\ref{cpw_coupler}), which is often used in quantum circuits to couple resonators to transmission lines. As in the single-line CPW case, we aim to compute the admittance matrix and the inductive energy. The procedure is a generalization of the one presented above for the single-line CPW and is largely based on the theory presented in \cite{Besedin_2018, Simon}. 

For a multi-line CPW with $n$ parallel and straight conductor lines ($n=2$ or $n=3$ in our case), a couple of equations similar to \eqref{phasor_solution} can be introduced for each conductor line. In array form
\begin{subequations}
\begin{align}
    \mathbf{V}(x) &= \mathbf{V}^+ e^{-jkx} + \mathbf{V}^- e^{jkx} \label{unovect} \\
    \mathbf{I}(x) &= \mathbf{I}^+ e^{-jkx} + \mathbf{I}^- e^{jkx} \label{duevect}
\end{align}
\label{phasor_solution_vect}
\end{subequations}
where  \(k = \omega/v \) as for the single line case. The arrays of forward and backward traveling wave coefficients $\mathbf{V}^\pm$ and $\mathbf{I}^\pm$ are related by 
\begin{equation}
\mathbf{V}^\pm = \pm \mathbf{Z} \mathbf{I}^\pm
\label{vipm}
\end{equation}
where the $n\times n$ characteristic impedance matrix \(\mathbf{Z}\) is given by 
\begin{equation}
    \mathbf{Z}=\sqrt{\mathbf{L}\mathbf{C}^{-1}}=\sqrt{\mathbf{C}^{-1}\mathbf{L}}
\end{equation}
with \( \mathbf{L} \) and \( \mathbf{C} \) the $n\times n$ matrices of inductances and capacitances per unit length, respectively. The latter satisfy the condition
\begin{equation}
\mathbf{L} \mathbf{C} = \frac{1}{v^2} \mathbf{I}_n
\label{LC}
\end{equation}
$\mathbf{I}_n$ being the $n\times n$ identity matrix and $v$ the propagation velocity given as for the single line case by $v=c/\sqrt{\varepsilon_\mathrm{eff}}$.

The most challenging aspect is the determination of the capacitance matrix, which can be obtained with conformal mapping techniques \cite{Besedin_2018, Simon} transforming the geometry into an equivalent parallel-plate capacitor configuration, from which the capacitances can be easily extracted. More details on this approach and its implementation in QuLTRA are given in Appendix \ref{conformal_mapping}.

 Being the central conductor, if present, grounded at both ends, the coupler is a 4-port device characterized by a $4\times4$ admittance matrix $\mathbf{Y}_\mathrm{cpl}(z)$ for both the $n=2$ and $n=3$ cases. The calculation  of the $i$-th column of $\mathbf{Y}_\mathrm{cpl}$ for a coupler of length $\ell$ proceeds with setting the voltage of the $i$-th port (located either at $x=0$ or $x=\ell$) equal to 1 V with all other ports grounded, then calculating $\mathbf{V}^\pm$ from \eqref{unovect}, $\mathbf{I}^\pm$ from \eqref{vipm}, and finally the corrent profiles $\mathbf{I}(x)$ and hence the port currents from \eqref{duevect}.

Similarly, the inductive energy $E_\mathrm{cpl}$ for mode $m$ is calculated as 
\begin{equation}
    E_\mathrm{cpl} = \frac{1}{2} \int_0^\ell \mathbf{I}(x)^\dagger \, \mathbf{L} \,\mathbf{I}(x) \, \mathrm{d}x
\end{equation}
where the array of current profiles is computed with a procedure similar to the one outlined above, but starting from the port voltages equal to the node voltages for mode $m$, and the inductance matrix \( \mathbf{L} \) is derived from the capacitance matrix \( \mathbf{C} \) using \eqref{LC}.
The integral can be evaluated analytically.

\section{Validation of the method through QuLTRA}
\label{sec:validation}
The proposed method was validated through QuLTRA, with its results compared to trusted references and other software packages. The validation follows two complementary approaches:
\begin{itemize}
    \item comparison with Ansys HFSS \cite{ansia} and pyEPR \cite{pyEPR}, which use full electromagnetic simulations;
    \item comparison with results from literature.
\end{itemize}
Additionally, a number of validation tests against QuCAT \cite{Gely_2020} have been performed for lumped-element circuits, always resulting in 
almost perfect agreement. Here we do not provide details of such comparisons, since lumped-element circuits are not the main focus of this paper. Anyhow, in the next section an application example that is simulated also with QuCAT is provided. More examples can be found in the QuLTRA GitHub repository.

The two points above are discussed in more details in the following subsections. 

\subsection{ Comparison with HFSS and pyEPR}
\begin{figure}
    \centering
    \begin{subfigure}[b]{0.45\textwidth}
        \centering
        \includegraphics[width=\textwidth]{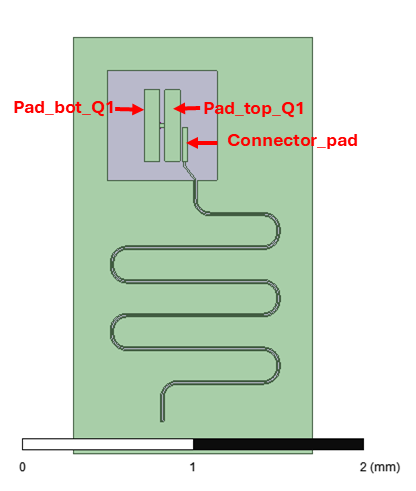}
        \caption{}
        \label{qubit_lambda_a}
    \end{subfigure}
    \hfill
    \begin{subfigure}[b]{0.45\textwidth}
        \centering
        \includegraphics[width=1\textwidth]{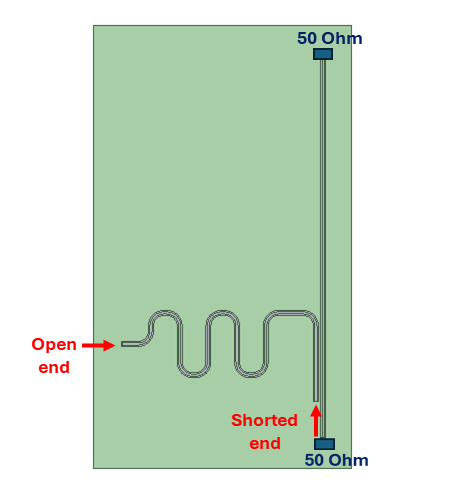}
        \caption{}
        \label{qubit_lambda_b}
    \end{subfigure}
    \caption{Test structures used for comparison between Ansys+pyEPR and QuLTRA. (a) Transmon qubit capacitively coupled to a $\lambda/2$ resonator taken from Qiskit Metal tutorials. (b) $\lambda/4$ resonator inductively coupled to a feedline terminated with 50-$\Omega$ loads at both ends. The red labels highlight the open/short terminations.}
\end{figure}

\begin{figure}
    \centering
    \includegraphics[width=0.9\linewidth]{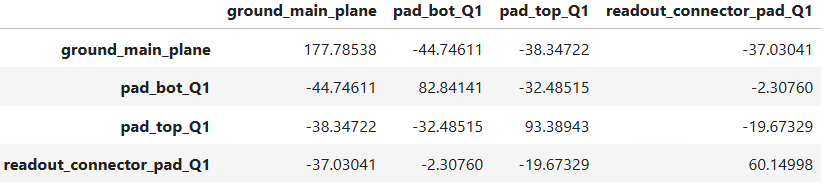}
    \caption{Capacitance matrix (in fF) of the transmon qubit used in the first test example extracted by Ansys Q3D capacitance extractor. The pad labels correspond to the ones indicated in Fig.\,\ref{qubit_lambda_a}.}
    \label{capacitance_matrix}
\end{figure}

The first test structure discussed here is a transmon qubit coupled to a half-wavelength ($\lambda/2$) readout resonator (Fig.\,\ref{qubit_lambda_a}). This setup is part of the Qiskit Metal tutorials~\cite{example} and a step-by-step analysis can be found in the QuLTRA GitHub repository. The capacitance matrix of the transmon qubit, extracted using Ansys capacitance extractor Q3D interfaced with Qiskit Metal, is reported in Fig.~\ref{capacitance_matrix}. The Josephson junction has a capacitance $C_j = 0$ and an inductance $L_j = 12.31\,$nH, while the line forming the $\lambda/2$ resonator is 6-mm long, with a width of 10\,\textmu m and a gap of 6\,\textmu m, and with the bottom end left open. In QuLTRA the resonator is represented by a 6-mm CPW line with characteristic impedence of $50\,\Omega$ as calculated according to \cite{Simon}.

The results obtained with HFSS+pyEPR and QuLTRA are compared in Table~\ref{tab:comparison}, which shows a general good agreement. The largest differences ($< 8 \%$) are in the anharmonicity and cross-Kerr values and to a minor extent in the qubit frequency. This is probably due to QuLTRA modeling the qubit through its capacitance matrix, as opposed to the full electromagnetic simulation in Ansys.

%Small differences appear in the quantities related to the qubit, such as its frequency, anharmonicity, and cross-Kerr. We attribute these discrepancies to convergence issues in Ansys; in fact, the deviation observed in the qubit frequency tends to vanish as the number of adaptive passes in the simulation increases. NON SO SE DIRLO, PERCHE UNO POTREBBE DIRE ALLORA LO FACEVI CONVERGERE DI PIU E IO GLI RISPONDO SI, MA DEVO ANDARE SUL SERVER CHE HA LINUX E QUINDI NON SI INTERFACCIA CON QISKIT PER FARE EPR?!
%
\begin{table}[h!]
  \centering
  \begin{tabular}{lcc}
    \toprule
    & Ansys HFSS results & QuLTRA results \\
    \midrule
    Qubit mode frequency (GHz) & 5.76 & 5.85 \\
    Resonator mode frequency (GHz) & 9.36 & 9.34 \\
    Qubit anharmonicity (MHz) & 348 & 321 \\
    Cross-Kerr (MHz) & 1.8 & 1.93 \\
    \bottomrule
  \end{tabular}
  \caption{Comparison between Ansys HFSS and QuLTRA results for the transmon–$\lambda/2$ resonator structure of Fig.\,\ref{qubit_lambda_a}.}
  \label{tab:comparison}
\end{table}

The second test structure is a quarter-wavelength ($\lambda/4$) resonator, 3.5-mm long, inductively coupled to a feedline terminated with 50-$\Omega$ resistances at both ends (Fig.\,\ref{qubit_lambda_b}). Both the feedline and the resonator line have a width of 15\,\textmu m and a gap of 10\,\textmu m. The coupler is a 3-line CPW with central grounded line of 15-\textmu m width.
Despite its simplicity, this structure is challenging to simulate with Ansys HFSS, especially for rather short coupler lengths, due to the computational time required to achieve convergence. 
 %To address this, we performed the simulations on a dedicated server, enforcing convergence criteria on the quality factor $Q$ rather than solely on the real part of the eigenfrequency. 
In QuLTRA the resonator is modeled with a CPW line with characteristic impedance of $51.6\,\Omega$ and length $\ell_R$, with one end left open and the other end connected to one port of the 3-line coupler of length $\ell$. The other ports of the coupler are connected to ground and to 50-$\Omega$ resistances (Fig.\,\ref{qubit_lambda_b}). It is $\ell_R + \ell = 3.5\,$mm, that is the total length of the resonator line in the layout. 
The coupler length $\ell$ is taken as the sum of the length of the straight part of the coupler plus half length of its curved section in the layout. 

\begin{figure}[h!]
    \centering
    \includegraphics[width=0.8\linewidth]{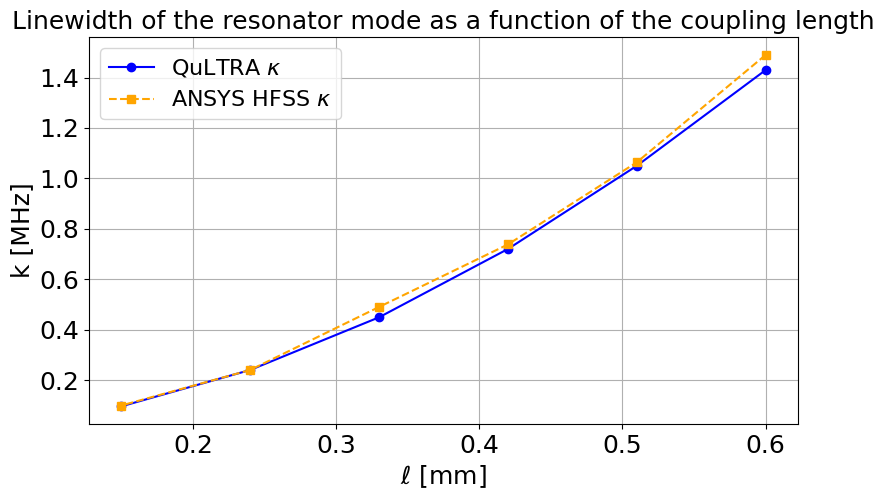}
    \caption{Linewidth of the $\lambda/4$ resonator coupled to a feedline as a function of the length of the coupler obtained with QuLTRA (solid line, blue) and Ansys HFSS (dashed line, orange).}
    \label{confronto}
\end{figure}

Fig.\,\ref{confronto} shows the values of the resonator linewidth $\kappa$ obtained with both QuLTRA and Ansys as a function of the coupler length $\ell$. The results are in very good agreement, in spite of large differences in the required computational time: QuLTRA completes the simulation in only a few seconds on a standard laptop equipped with an AMD Ryzen 7 7730U (8 cores) and 16 GB RAM running Windows 11, whereas the equivalent Ansys HFSS simulation requires nearly one hour on a high-performance server (HPE ProLiant DL560 Gen10) featuring 96 physical Intel Xeon Gold 6252N cores and 512 GB RAM.
\subsection{Comparison with results from literature}
\begin{figure}
    \centering
        \includegraphics[width=0.8\textwidth]{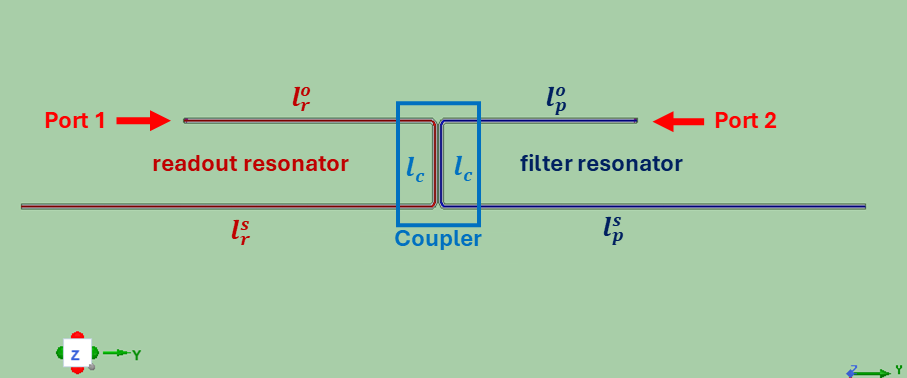}
        \caption{Schematic layout of the notch Purcell filter for fast qubit readout proposed in \cite{Spring_2025}, where a  $\lambda/4$ readout resonator and a $\lambda/4$ filter resonator are coupled through a multi-line coupler. In the complete circuit port 1 and port 2 are capacitively connected to the qubit and to the readout transmission line, respectively, while the other two line ends are grounded.}
        \label{zurigo_structure}
\end{figure}

In order to further validate the capability of the method to analyze circuits containing distributed element components such as transmission lines and couplers, we consider here the structure proposed in \cite{Spring_2025}. It consists of a filtering circuit for fast qubit readout, based on a $\lambda/4$ CPW readout resonator and a $\lambda/4$ CPW filter resonator coupled to each other through a multi-line coupler in such a way to produce a compact notch filter to effectively eliminate the Purcell decay of the qubit. 

A schematic layout is shown in Fig.\,\ref{zurigo_structure}.
The characteristic impedance of the resonators is approximately $66\,\Omega$, with the line width of 5\,\textmu m
and the gaps of 7.5\,\textmu m. In the coupler section there is an additional central grounded line of 5.5\,\textmu m width.
Each resonator is formed by three sections: the first one extending from the corresponding port (port 1 or port 2) to the coupler with length $\ell_{r/p}^0$, the second one shared with the coupler with length $\ell_c$, and the third one from the coupler to ground with length $\ell_{r/p}^s$ (see Fig.\,\ref{zurigo_structure}). The precise values are $\ell_r^0 = 981.5\,$\textmu m, $\ell_r^s = 1624.5\,$\textmu m, $\ell_p^0 = 766.5\,$\textmu m, $\ell_p^s = 1666.5\,$\textmu m and $\ell_c = 335\,$\textmu m.
The total lengths of the two resonators are $\ell_{r}^0+\ell_r^s+\ell_c = 2.94\,$mm and $\ell_{p}^0+\ell_p^s+\ell_c = 2.77\,$mm, respectively.

\begin{figure}
    \centering
    \includegraphics[width=0.8\textwidth]{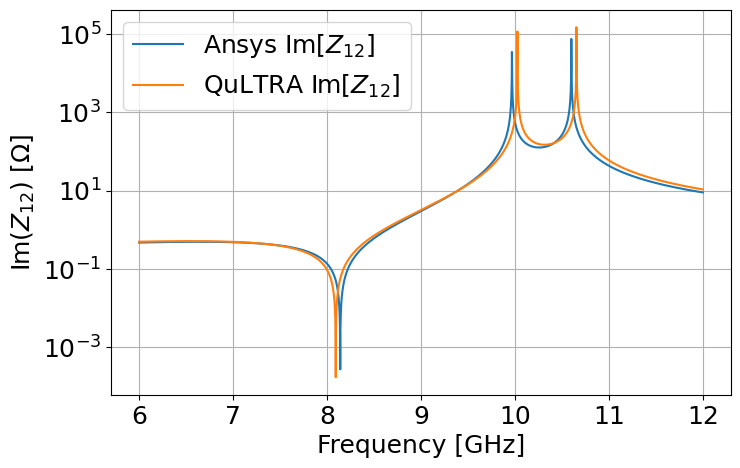}
       \caption{Imaginary part of the element $Z_{12}$ of the impedance matrix of the 2-port structure of Fig.\,\ref{zurigo_structure}, closely inspired by the design reported in \cite{Spring_2025}, calculated with QuLTRA (orange) and Ansys HFSS (blue).}
        \label{Z12}
\end{figure}

As a first step, we used QuLTRA to compute the 2-port impedance matrix of the filtering circuit. The same calculation was performed using also Ansys HFSS for comparison. The plot of the resulting $Z_{12}$ element is shown in Fig.\,\ref{Z12}, which shows a very good agreement. It exhibits a notch and two peaks corresponding to the resonance frequencies of the two resonators, closely resembling the one presented in the original paper.

\begin{figure}
        \centering
        \includegraphics[width=0.8\textwidth]{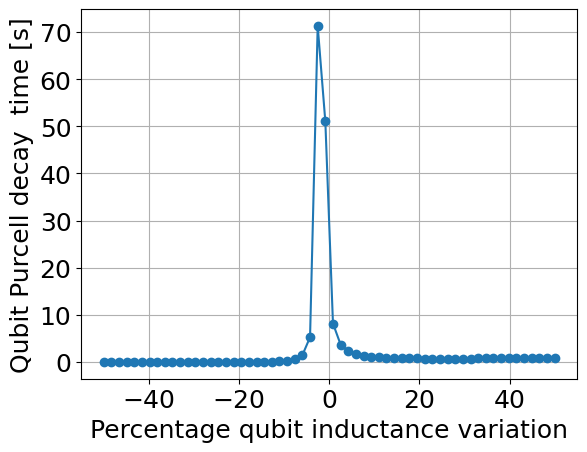}
        \caption{Purcell decay time of a qubit using the readout filtering structure of Fig.\,\ref{zurigo_structure} as a function of the relative variation of the qubit inductance with respect to its nominal value, which corresponds to the qubit frequency being perfectly tuned to the notch frequency of the filter.} 
        \label{purcell}
\end{figure}

To complete the validation, we analyzed the full circuit obtained by connecting the above structure to a qubit via port 1 through a coupling capacitor $C_{g} = 4\,$fF and to a readout line terminated on a 50-$\Omega$ load via port 2 through a capacitance $C_{k} = 10\,$fF. The transmon qubit is modeled with a total shunt capacitance $C_{q} = 55.5\,$fF and a parallel Josephson junction with nominal inductance $L_{j} = 6.6\,$nH. The purpose is to 
evaluate the Purcell decay time, calculated as the inverse linewidth of the qubit mode, $1/\kappa_q$. By slightly varying the qubit inductance, and thus the qubit frequency, around its nominal value, it is possible to verify the sensitivity of the Purcell decay time with respect to variations of the qubit inductance. The result is shown in Fig.~\ref{purcell}. In the nominal condition, when the qubit frequency is perfectly tuned to the notch frequency, the Purcell decay time is extremely high, but it rapidly deteriorates even for small deviations of the inductance value. 
\section{Further validation examples}
\label{sec:examples}
In this section further examples of application are considered where the capability to handle directly distributed components turns out to be particularly useful.
\subsection{Qubit and resonator in multimode ultra-strong coupling regime}
The first example is a system formed by a transmon qubit capacitively coupled to a $\lambda/4$ resonator and operated in the multimode ultra-strong coupling regime~\cite{Bosman_2017}.
In this regime the transmon is coupled to multiple modes of the resonator and it is a priori unclear how many modes it is necessary to consider for a realistic modeling of the system. This configuration is also analyzed in the QuCAT article~\cite{Gely_2020}, allowing a direct comparison between QuLTRA and QuCAT.
The qubit capacitance is $C_j = 5.13$ fF, the Josephson inductance $L_j = 9$ nH  and the coupling capacitance $C_g = 40.3$ fF. Since QuCAT only handles lumped elements, in \cite{Gely_2020} the multimode resonator is modeled using Foster's reactance theorem \cite{Foster} as a series of parallel LC resonators, each corresponding to a resonator mode. This approximation becomes more and more accurate as the number of LC sections is increased. For this reason, to accurately determine the qubit mode frequency and anharmonicity, at least 10 LC sections are required in QuCAT. This increases the number of circuit nodes and consequently the computational time. 
In QuLTRA, instead, since the resonator is treated as a genuine CPW line, the qubit mode frequency and anharmonicity, resulting in 8.02 GHz and 352 MHz, respectively, are directly determined without further approximations. This not only saves time, but also removes the need to worry about the number of required modes. 

\begin{figure}
    \centering
    \includegraphics[width=0.7\linewidth]{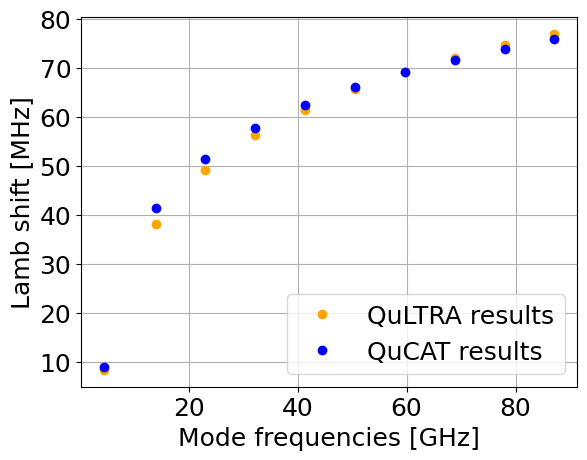}
    \caption{Lamb shift of the qubit frequency in the ultra-strongly coupled qubit/resonator system as a function of the highest mode frequency of the resonator considered in the calculation.}
    \label{lamb_shift}
\end{figure}

Finally, the Lamb shift is calculated using \eqref{chi_b} as a function of the number of modes (Fig.\,\ref{lamb_shift}) and, in agreement with QuCAT, its value tends to grow and slowly saturate with the number of modes, even though the qubit and the resonator are well separated in frequency.
\begin{figure}[h!]
    \centering
    \begin{subfigure}[b]{0.45\textwidth}
        \centering
        \includegraphics[width=1.1\textwidth]{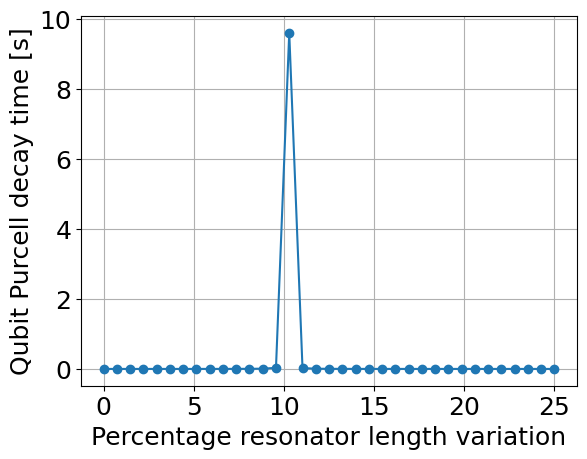}
        \caption{}
        \label{purcell_pole}
    \end{subfigure}
    \hfill
    \begin{subfigure}[b]{0.5\textwidth}
        \centering
        \includegraphics[width=1.2\textwidth]{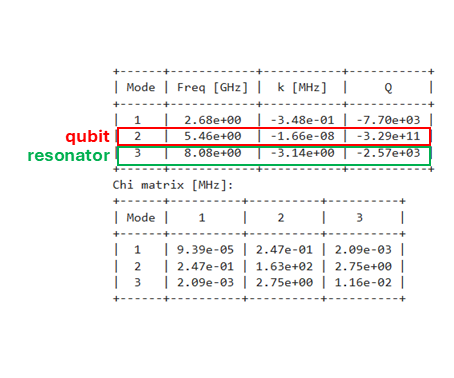}
        \caption{}
        \label{purcell_result}
    \end{subfigure}
    \caption{QuLTRA simulation results of the notch Purcell filter exploiting higher frequency poles of a $\lambda/4$ readout resonator. (a) Purcell decay time vs.\,\,variation of the resonator length relative to its nominal value of 9.8 mm. (b) Top: mode frequencies, linewidths and quality factors. Bottom: anharmonicities and cross-Kerr matrix. Calculation performed with the optimal resonator length resulting from (a). Mode 2 (red) is the qubit mode, with 5.46-GHz frequency and 163-MHz anaharmonicity. Mode 3 is the readout mode (green), corresponding to the second resonator mode at 8.08 GHz.}
\end{figure}
\subsection{Notch Purcell filter leveraging higher frequency poles of a $\lambda/4$ readout resonator}
In this second example we consider a transmon qubit with a frequency of 6 GHz and an anharmonicity of 200 MHz, coupled to a $\lambda/4$ readout resonator through a coupling capacitance $C_g=20$ fF. The resonator is 9.8-mm long, has a characteristic impedance of 50 $\Omega$ and is coupled to the 50-$\Omega$ readout line via a capacitance $C_k=12$ fF.
Due to the distributed nature of the resonator, its resonance frequencies, which are close to the zeros of its admittance, alternate with the poles of the same admittance. An intrinsic notch Purcell filter can then be obtained 
by aligning the qubit mode frequency with one of such poles, resulting in the qubit being effectively shorted at the qubit frequency, thus preventing energy leakage into the readout line. Note that, in order to place a resonator pole at the qubit frequency, the fundamental mode of the resonator must lie below the qubit frequency. Since normally the readout frequency above the qubit frequency is the preferred condition, the second resonator mode can be used for readout. 

The analysis and sizing of such a structure can be conveniently carried out using QuLTRA. 
A slight change in the nominal resonator length allows to fine tune the resonator pole to the qubit frequency, leading to a peak in the Purcell decay time, $1/\kappa_q$, as shown in Fig.\,\ref{purcell_pole}. At the optimal resonator length, the qubit shows a dissipation rate $\kappa_q/2\pi = 0.0166\,$Hz (Fig.\,\ref{purcell_result}), which corresponds to a Purcell decay time of 9.59 s. The linewidth of the readout mode (mode 3, the second resonance mode) is $\kappa_r/2\pi\ = 3.14\,$MHz, and is close to the dispersive frequency shift, $\chi_{qr}/2\pi = 2.75\,$MHz, as required to optimize the signal to noise ratio. 
\subsection{Multiplexed qubit readout scheme}
\begin{figure}
    \centering
    \includegraphics[width=0.8\linewidth]{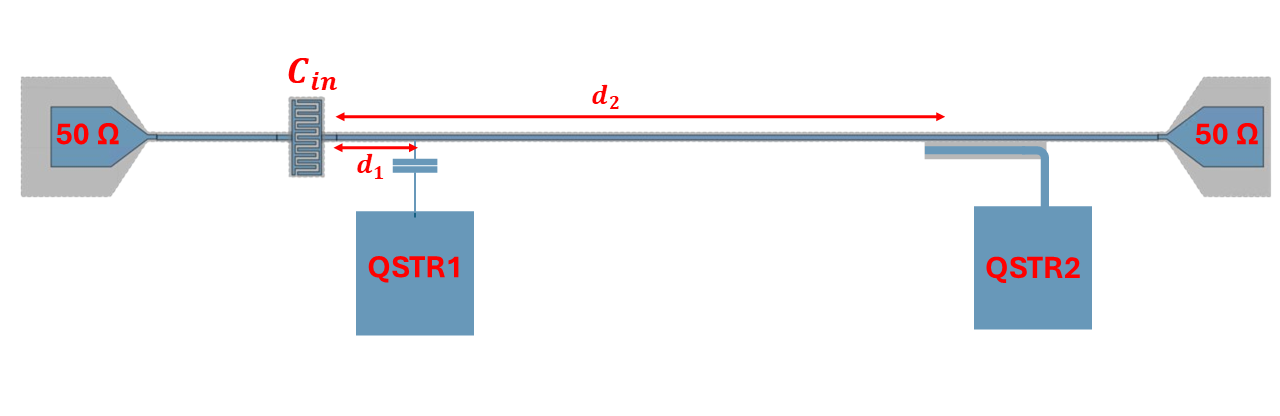}
    \caption{Schematic representation of the multiplexed qubit readout circuit simulated with QuLTRA. The readout CPW line is terminated at its left end with a capacitance $C_\mathrm{in} = 42.7\,$fF in series with a 50-$\Omega$ load, while  its right end is impedance matched. QSTR1 is a subcircuit formed by qubit + readout resonator + Purcell filter resonator and is capacitively coupled to the line. QSTR2 is formed only by qubit + readout resonator with no Purcell filter and is inductively coupled to the line through a CPW coupler of the type discussed in the previous sections. $d_1$ and $d_2$ are the distances of the two coupling points from capacitance $C_\mathrm{in}$.}
    \label{feedline}
\end{figure}
As third example, we consider a common configuration for the readout of multiple qubits \cite{Heinsoo}, where several qubit structures, each comprising a qubit, a readout resonator, and possibly a Purcell filter, are all connected to a shared CPW readout line (see Fig.\,\ref{feedline}, where two qubit structures, QSTR1 and QSTR2, are considered). It is assumed that the readout line has 50-$\Omega$ characteristic impedance and is connected at one end to a 50-$\Omega$ load and at the other end to an input capacitance $C_\mathrm{in}$ in series with a 50-$\Omega$ load. 
Since the line is not impedance matched at both ends, standing waves are formed along the line, causing the linewidths of the readout resonators to depend on their coupling position along the line. 
Specifically, for a readout resonator with frequency $f_r$ capacitively coupled to the line,
linewidth maxima are expected to occur at distances from capacitance $C_\mathrm{in}$ equal to integer multiples of $\lambda_r/2$, with $\lambda_r = v / f_r$, $v$ being the propagation velocity on the line \cite{Heinsoo}, while for the case of inductive coupling linewidth maxima are expected at integer multiples of $\lambda_r/4$.

\begin{figure}[h!]
    \centering
    \includegraphics[width=0.8\linewidth]{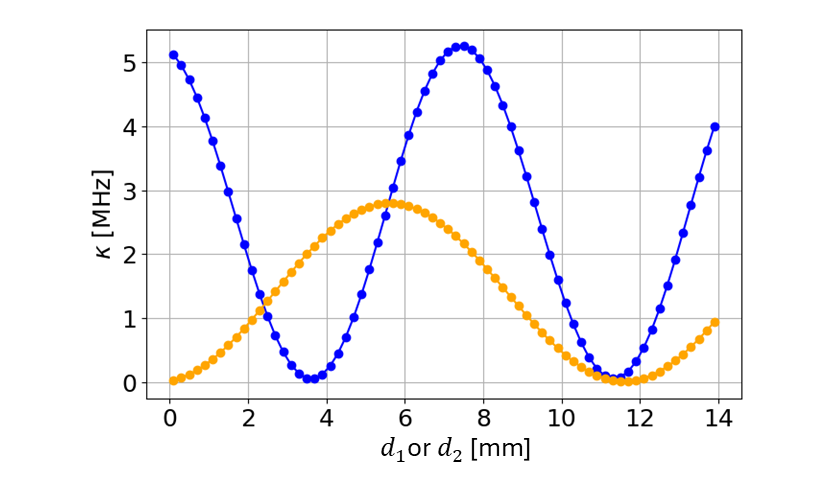}
    \caption{Linewidth of the readout resonator in QSTR1 (blue) and QSTR2 (orange) as a function of the distance from \(C_\mathrm{in}\) of the respective coupling point to the line, $d_1$ and $d_2$, defined in Fig.\,\ref{feedline}.}

    \label{coupling_type}
\end{figure}

The analysis of this type of circuit can be easily carried out with QuLTRA. In the example presented here, two qubit structures are considered, as shown in Fig.\,\ref{feedline}. The first one, referred to as QSTR1, consists of a qubit with frequency close to 5.5 GHz, a $\lambda/4$ readout resonator with near to 7.5 GHz frequency and a second $\lambda/4$ resonator acting as Purcell filter, capacitively coupled to the readout line. The second one, called QSTR2, consists of a 4-GHz qubit and a 5-GHz $\lambda/4$ readout resonator inductively coupled to the readout line through a CPW coupler. No Purcell filter is present in QSTR2. The parameter values for QSTR1 and QSTR2 are shown in Table~\ref{tab:first_structure} and \ref{tab:second_structure}, respectively. Fig.\,\ref{coupling_type} shows the calculated linewidths of the two readout resonators as a function of the distances $d_1$ and $d_2$ from $C_\mathrm{in}$. The results are consistent with the previous discussion: for QSTR1, which is capacitively coupled to the line, a linewidth maximum occurs at approximately 7.6 mm, corresponding to half wavelength at the readout resonator frequency, while for the inductively coupled QSTR2 a linewidth maximum occurs at about 5.85 mm, that is a quarter wavelength at the readout resonator frequency.

\begin{figure}[ht]
    \centering
    \includegraphics[width=0.9\linewidth]{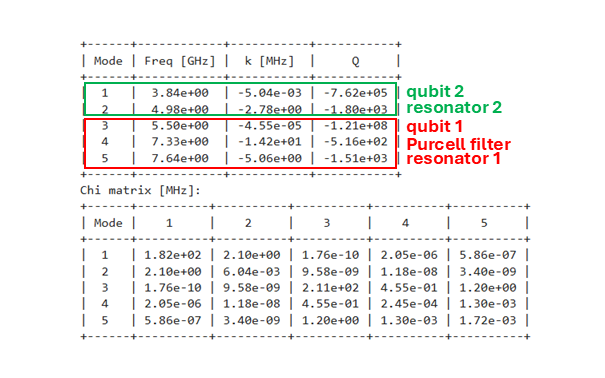}
    \caption{Table of complete results of QuLTRA simulations obtained for the two qubit structures QSTR1 and QSTR2 located at the distances $d_1$ and $d_2$ from \(C_\mathrm{in}\) corresponding to maximun coupling with the line (see Fig.\,\ref{coupling_type}). Top: mode frequencies, linewidths and quality factors. Bottom: anharmonicities and cross-Kerr matrix. Colors and labels help to identify the qubit and resonator modes: red and green for the components of QSTR1 and QSTR2, respectively.}

    \label{complex}
\end{figure}

Fig.\,\ref{complex} reports the complete set of QuLTRA results for the circuit with both qubit structures located at the distances $d_1$ and $d_2$ corresponding to the peak resonator linewidths. The different colors and labels help to identify the different qubit and resonator modes. 

This example demonstrates the usefulness of the proposed method in analysing complex configurations including several qubits and resonators, some of which are CPW based, that would be hard to simulate with full electromagnetic solvers.

\begin{table}[ht]
\centering
\begin{tabular}{lc}
\toprule
Component & Value \\
\midrule
Qubit capacitance & 85 fF \\
Qubit inductance & 9.2 nH \\
Qubit-resonator coupling capacitance & 6 fF \\
Readout resonator length & 3.8 mm \\
Filter length & 3.8 mm \\
Filter-readout coupling capacitance & 12 fF \\
Coupling capacitance to feedline & 20 fF \\
\bottomrule
\end{tabular}
\caption{Component values for the qubit + readout resonator + Purcell filter structure QSTR1 of Fig.\,\ref{feedline}. The readout resonator and the Purcell filter are based on CPW lines, whose lengths are reported in the table.}
\label{tab:first_structure}
\end{table}

\begin{table}[ht]
\centering
\begin{tabular}{lc}
\toprule
Component & Value \\
\midrule
Qubit capacitance & 97 fF \\
Qubit inductance & 16.3 nH \\
Qubit-resonator coupling capacitance & 8 fF \\
Readout resonator length & 5.85 mm \\
Coupler length &  0.7 mm \\
\bottomrule
\end{tabular}
\caption{Component values for the qubit + readout resonator structure QSTR2 of Fig.\,\ref{feedline}. The readout resonator is based on a CPW line of length indicated in the table. The length of the CPW coupler used to connect the structure to the line is also reported in the table.}
\label{tab:second_structure}
\end{table}

\section{Conclusion}
\label{sec:conclusions}
We have presented a general method for analyzing superconducting quantum circuits combining lumped and distributed components. The method has been validated through its implementation in a dedicated package called QuLTRA. By directly modeling distributed elements such as CPW resonators and couplers, the method eliminates the need for heavy lumped-element discretization, significantly reducing the number of circuit nodes. Validation against electromagnetic simulations, existing software packages and experimental designs from the literature shows that the approach achieves excellent accuracy across a wide range of circuit topologies. The discussed application examples, including Purcell protected readout, multi-mode ultra-strong coupling, and multiplexed qubit readout, highlight the tool versatility in addressing engineering optimization studies in circuit QED. 

We stress that, even if the examples presented here involve only ideal (i.e.~lossless and straight) CPW transmission lines and multi-line couplers, the method lends itself to ample generalizations. Indeed, any circuit that in its linearized version can be interpreted as a network of multi-port components, each described by an admittance matrix, can in principle be analysed with the proposed method.

\clearpage
\appendix
\section{Calculation of the capacitance matrix of the CPW coupler}
\label{conformal_mapping}
The calculation exploits the conformal mapping technique to remap a multi-conductor transmission line into an equivalent parallel-plate capacitor \cite{Besedin_2018, wang2015conformalmappingmultipleterminals}.
\begin{figure}[h!]
    \centering
    \begin{subfigure}[b]{0.8\textwidth}
        \centering
        \includegraphics[width=0.9\textwidth]{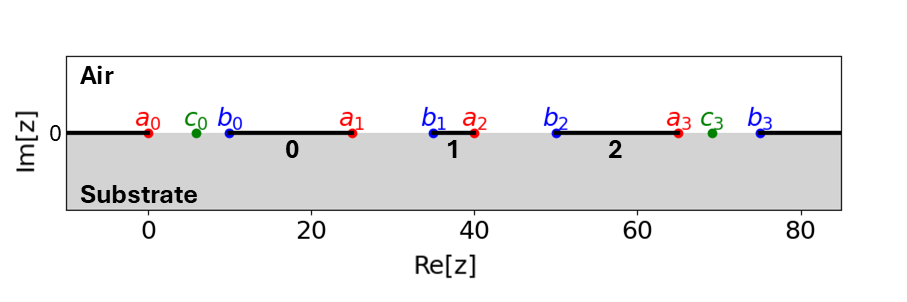}
        \caption{}
        \label{cpw3}
    \end{subfigure}
    \hfill
    \begin{subfigure}[b]{0.8\textwidth}
        \centering
        \includegraphics[width=0.9\textwidth]{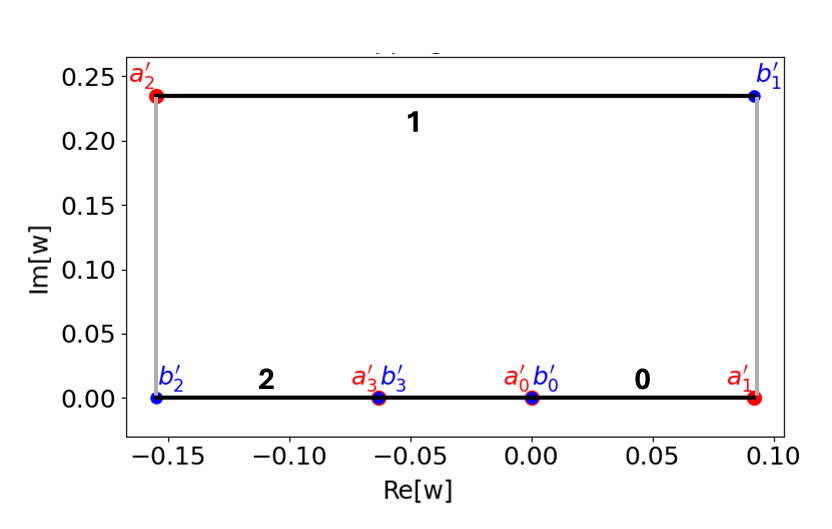}
        \caption{}
        \label{map}
    \end{subfigure}
    \caption{(a) Representation in the complex $z$ plane of the cross-section of a 3-line CPW structure. Conductors are indicated by solid-line segments: lines are marked with 0, 1 and 2, while the lateral conductors belong to the ground planes. All edge points are located on the real axis, the right edges \(a_\mathrm{i}\) are shown in red, the left edges \(b_\mathrm{i}\) in blue. The two half-planes corresponding to $\mathrm{Im}[z]>0$ and $\mathrm{Im}[z]<0$ are mapped by the transformations indicated in the text onto rectangular domains in the complex $w$ plane. (b) Result of the transformation in the $w$ plane applied for the calculation of the column (row) corresponding to line 1 of the capacitance matrix. Points \(c_0\) and \(c_3\) indicated in green in (a) are functional to such transformation (see text).
}
    \label{mapping}
\end{figure}
The cross-section of a 3-line CPW is shown in Fig.~\ref{cpw3}, represented in the complex domain $z$ where the substrate-air interface is located on the real axis. The half-plane corresponding to negative imaginary part (grey area) is the substrate. The metal lines have zero thickness and are represented by solid-line segments. The external segments belong to the lateral ground plane. 
The right and left edges of the conductor lines are indicated by 
$a_{i}$ and $b_{i}$, respectively. The structure is supposed to be uniform, that is the permittivity is constant in each half-plane. 

The capacitance matrix is calculated by columns. For the calculation of the $m$-th column (or row, since the matrix is symmetric), with $m=0,...,n-1$ and $n$ the number of conductor lines, the $m$-th conductor is held at a nonzero potential $\phi_{m}$, while all other conductors are grounded. By applying a suitable form of Schwarz–Christoffel transformation \cite{Besedin_2018}, each half-plane of the $z$ domain is mapped into a rectangular domain of the complex $w$ plane (see Fig.~\ref{map}), where the conductor at potential $\phi_{m}$ is mapped onto the upper boundary of the rectangle, while all the other grounded conductors are aligned on the lower boundary. The Schwarz–Christoffel transformation used in this context is given by
\begin{equation}
w(z) = \int_{0}^{z}
\frac{
  \displaystyle\prod_{j\in\{0,\dots,n\}\setminus\{m,m+1\}}(z^\prime - c_j)
}{
  \displaystyle\prod_{k=0}^{n}(z^\prime - a_k)^{\!1/2}
  \;\displaystyle\prod_{l=0}^{n}(z^\prime - b_l)^{\!1/2}
}
\,\mathrm{d}z^\prime
\label{w(z)}
\end{equation}
The points $c_j$, with $j\in\{0,\dots,n\}\setminus\{m,m+1\}$, are located in the gaps between two grounded conductors and are calculated by imposing that the integral over each of such gaps is zero, thus mapping all the grounded metal lines onto an uninterrupted line corresponding to the lower boundary of the rectangle (see Fig.~\ref{map}).

The interelectrode capacitances can then be easily calculated using the parallel plate capacitor formula. By defining $a_i^\prime = w(a_i)$ and $b_i^\prime = w(b_i)$,
the elements of the $m$‑th column of the capacitance matrix can be written as
\begin{equation}
    C_{im}=(\varepsilon_r+1)\varepsilon_0\frac{\operatorname{Re}(b'_\mathrm{i})-\operatorname{Re}(a'_\mathrm{i+1})}{\operatorname{Im}(b'_\mathrm{m})
}
\end{equation}
Notice that the off-diagonal elements are negative, while the diagonal elements are equal to the positive sum of the off-diagonal elements of the same row. 

\bibliographystyle{unsrt}\bibliography{citation}
\end{document}